# Density of States (Gate) - Controlled Andreev Molecule and Sensor

Xiaofan Shi[1,4,†], Ziwei Dou[1,†,*], Guoan Li[1,4,†], Dong Pan[2,†], Yuxiao Song[1,4,†], Anqi Wang[1], Zhiyuan Zhang[1,4], Xingchen Guo[1,4], Xiao Deng[1,4], Ruixuan Zhang[1,4], Liangqian Xu[1,4], Xiao Chen[1,4], Yupeng Li[5], Bingbing Tong[1], Xiaohui Song[1], Zhaozheng Lyu[1], Peiling Li[1], Fanming Qu[1,4,6,7], Guangtong Liu[1,6,7], Jianhua Zhao[2,3,*], Li Lu[1,4,6,7,*], Jie Shen[1,6,*]

[1]Beijing National Laboratory for Condensed Matter Physics and Institute of Physics, Chinese Academy of Sciences, Beijing 100190, China

[2]State Key Laboratory of Semiconductor Physics and Chip Technologies, Institute of Semiconductors, Chinese Academy of Sciences, Beijing 100083, China

[3]National Key Laboratory of Spintronics, Hangzhou International Innovation Institute, Beihang University, Hangzhou 311115, China

[4]University of Chinese Academy of Sciences, Beijing 100049, China

[5]Hangzhou Key Laboratory of Quantum Matter, School of Physics, Hangzhou Normal University, Hangzhou 311121, China

[6]Songshan Lake Materials Laboratory, Dongguan, Guangdong 523808, China

[7]Hefei National Laboratory, Hefei 230088, China

* Corresponding authors. E-mails: ziweidou@iphy.ac.cn, jhzhao@semi.ac.cn, lilu@iphy.ac.cn, shenjie@iphy.ac.cn.

†These authors contributed equally to this work.

# Abstract

Topological quantum computing typically relies on topological Andreev bound states (ABSs) engineered in hybrid superconductor-semiconductor devices, where gate control offers key advantages. While strong Zeeman fields can induce such states, an alternative approach emerges through Andreev molecules - closely spaced, coupled ABSs, also key building-block for Kitaev chain - that enable topological behavior without high magnetic fields. However, existing Andreev molecules are controlled via magnetic flux in superconducting loops, limiting scalability. Here, we introduce a gate-controlled Andreev molecule, where electrostatic tuning of the density of states in one site nonlocally enhances the critical current of another. This eliminates superconducting loops, offering superior tunability, scalability, and sensitivity. We further extend such an Andreev molecule to a multi-site Kitaev chain, and a noninvasive sensor resolving single-Cooper-pair charge for parity readout. This platform bridges the gap between scalable ABS engineering and high-sensitivity quantum sensing, advancing the development for constructing and parity-readout in topological ABSs and long Kitaev chains towards topological qubits.

# Main text

The pursuit of large-scale fault-tolerant quantum computer has focused intensely on topological quantum computing, which braids non-Abelian anyons in topological superconductivity[1,2]. While intrinsic topological superconductors remain challenging to realize, hybrid systems combining conventional superconductors and normal components have emerged as promising platforms to construct such non-Abelian states. Among the large family of candidates using topological materials[3,4] and ferromagnetic materials[5,6], hybrid Josephson junctions using semiconducting nanowires with strong spin-orbit coupling (SOC)[7,8] offer unparalleled advantages in device control and scalability - essential requirements for practical topological quantum computing.

Josephson junctions (JJs) - the fundamental building blocks of superconducting circuits - take on new significance in this context by engineering the hybrid architectures towards topological qubit. In conventional JJs, the supercurrent flowing between two superconducting contacts is transmitted by tunneling of Cooper pairs through a thin layer of insulator. However, in the hybrid systems where the insulator is replaced by materials with finite density of states (DOS) such as the semiconducting nanowires, the supercurrent is mediated by the phase-coherent Andreev bound states (ABSs) formed in between[9]. The ABSs can be made topologically nontrivial if the band in the semiconducting nanowire is inverted by inducing both strong SOC and large Zeeman energy correlated with high magnetic field[7,8], or if the proximitized material itself possesses non-trivial topology[5,6]. Recently, engineering topological ABSs in highly controlled semiconducting hybrid systems without complications such as requiring Zeeman field or topological materials, has been explored via the coherent couplings of more than one ABSs (JJs), which is termed as Andreev molecule[10-15]. When separated by a distance shorter than the superconducting coherent length, the wavefunctions of ABSs in the neighboring JJs of the Andreev molecule may overlap. Due to the subsequent hybridization, by phase biasing one JJ, the ABSs of the other JJ of the Andreev molecule may exhibit time-reversal symmetry breaking (TRSB), a direct precursor to the emergence of topological superconductivity[16-20]. Furthermore, scaling the systems into multiple JJs, the ABSs of the Andreev molecule can be further engineered to acquire non-trivial topology via controlling the phase biases of each junction collectively[21-26]. Most remarkably, the network of coupled ABSs has intimate relation with the recently realized Kitaev chain, where the longer chain with multiple sites is proposed to enhance topological protection and rapidly rises as a promising protocol to implement topological qubit[27-30].

Besides its construction, the reliable readout of parity states in topological qubits also presents one of the most challenging tasks in quantum computation. Unlike conventional qubits that rely on fragile quantum superpositions, topological qubits encode information in the parity of non-local Majorana zero modes (MZMs) with the braiding operation, offering inherent protection against local decoherence[8]. However, the non-local nature at the same time makes their parity readout notoriously difficult for conventional measurement approaches without disturbing the system. One effective method is to employ elaborate interference loops and external charge sensors[34], which also complicates device architecture and introduces new noise sources. This inherent trade-off between topological protection and measurability highlights the critical need for compact, non-invasive sensing architectures that preserve the qubit's topological protection while enabling high-fidelity readout, and also should be particularly compatible with the rapidly evolving field of gate-controlled Andreev molecule and Kitaev Chain[27-32].

Now, let us consider the most basic form of a typical Andreev molecule with only two semiconducting JJs (defined as $JJ_1$ and $JJ_S$) separated by a common superconductor in Fig. 1a. Besides simple JJs, each site can also be configured as more complicated devices. When the length of the common superconductor $L$ is much longer than the coherence length $l_\xi$, the ABSs of the two sites are independent (Fig. 1b, dashed lines). When $L$ is comparable with or shorter than $l_\xi$, the ABSs of the two sites will hybridize at the level crossings (Fig. 1b, solid lines), and by tuning the ABSs of one site, the ABSs of the other will be modified non-locally. Conventionally, the Andreev molecule is implemented in the phase (flux)-controlled scheme (named as "Type I" Andreev molecule, Fig. 1c), where the phases of the each site $\varphi_1$ and $\varphi_S$ in the molecule ABS spectrum $E(\varphi_1,\varphi_S)$[11-13,15,16,19] are tuned via the magnetic flux defined by the superconducting loops. By tuning of $\varphi_1$ of one site, the ABSs of the other site $E(\varphi_S)$ will also be modulated nonlocally and may exhibit a TRSB anomalous phase shift[11-13,16-19,31] (Fig. 1d).

In this work, we identify a new scheme to control the Andreev molecule not by phase via magnetic flux in the external loop but by the density of states (DOS) via electric gating of each site in Fig. 1e (named as "Type II" Andreev molecule). In particular, we find that the critical current of one site ($JJ_S$) has a nonlocal and sensitive response to the DOS of the other ($JJ_1$), even if the critical current of the latter is significantly smaller than the first (Fig. 1f). We also numerically reproduce such DOS-controlled molecule effects, and identify its mechanism as the ABSs wavefunction hybridization and transmission, both of which are gate tunable. Furthermore, with the benefit of electrical gating, such nonlocal DOS control of the Andreev molecule can be developed into several more complex configurations, forming an extended Andreev molecule for multi-JJ chain and an Andreev molecule sensor enabling highly sensitive and noninvasive readout even for single Cooper pair charge states. Those configurations are highly relevant in building and sensing the JJ network for advanced Kitaev chain systems and the future topological qubits[27-30].

The new type DOS (gate)-controlled Andreev molecule has several advantages in comparison to the conventional phase-controlled scheme: First, the electrical gating in DOS (gate)-controlled Andreev molecule can independently control each site of the molecule, which is much more convenient and well-developed than the independent control of the flux in each loop for the phase (flux)-controlled Andreev molecules. Second, the DOS-controlled Andreev molecule does not require adding more superconducting loops with the increasing number of the sites, which facilitates further scaling up of the molecule to a complex network essential for engineering topologically nontrivial systems or topological qubit. Third, the electrical gating can be used to flexibly configure the molecule sites as the extended Andreev molecule chain or as the highly sensitive charge/parity sensor. Indeed, the DOS-controlled Andreev molecule here has close similarity to the recently demonstrated Kitaev chain[27-30]. In particular, the understanding of how the Andreev molecule responds nonlocally to the gate of each site is beneficial for the fine gate-tuning of the longer Kitaev chain towards topologically protected Majorana zero modes (MZMs) and topological qubits, as well as for the sensitive and noninvasive charge/parity readout of the topological Majorana states via the coupled Andreev molecule states[10,32,33].

**Andreev Molecule Setup and Measurement**

We fabricated the Andreev molecule device by a single InAs nanowire with epitaxial Al film (Fig. 2a). one site (right site) is a simple Josephson junction $JJ_S$ with the Al film entirely removed, whose DOS is controlled by SG. The other site (left site) has two $JJ_{1,2}$ (bare InAs nanowire) connecting in series by a small superconducting island (Al-InAs section). By tuning each junction with the tunnel gates TG1,2, the left site can be configured either as a single JJ, a chain of JJs, or a single Cooper pair transistor (CPT), thus forming a simple DOS-controlled Andreev molecule, an extended Andreev molecule with multiple sites, or an Andreev molecule sensor for single Cooper pair tunneling, respectively. If each site is further configured as a quantum dot, the DOS (gate)-controlled Andreev molecule may function as two-site or even three-site Kitaev chain[27-30]. The Al section between $JJ_1$ and $JJ_S$ is defined by electron-beam lithography followed by metal deposition (see Method). The length of this shared superconducting section $L \approx 300$ nm which is smaller than the coherence length of Al at T $\approx$ 10 mK.

In order to simultaneously read out the critical current $I_C$ of each site, we further connect the ends of two sites together to form a superconducting interference device (SQUID) and measure the total critical current of the SQUID versus the enclosed flux around zero field at different DOS controlled by gatings (Fig. 2b). However, it should be noted that the loop here is not necessary for the formation of this DOS-controlled Andreev molecule, but serves exclusively as a compact way to simultaneous measure the critical current of each site. Also, to boost measurement efficiency, this loop readout method is further combined with the fast counter measurement technique[36-40] which applies repeated fast dc current pulses and detects $I_C$ using a digital counter. Such technique obtains the same $I_C$ as the conventional lock-in technique but with much faster speed, and is widely accepted in supercurrent measurements (see Supplementary Information Sec. 1). Finally, we also emphasize that fundamentally different from the type I phase (flux)-controlled Andreev molecule[11-13,15,16,19], where the molecule is controlled by the flux through the SQUID loop and thus the number of SQUID loops is scaled up with the number of the sites in the molecule, the control of type II DOS (gate)-controlled Andreev molecule is realized by the local gates and the single SQUID loop described above is used only for compact readout of multiple sites. Such a superconducting loop may also be useful for the future charge/parity readout of the topological qubits[34,35].

**DOS (gate)-controlled Andreev molecule**

We first consider the most basic configuration, in which the single $JJ_1$ and $JJ_S$ form the Andreev molecule. SG and TG2 is set such that $I_{C,S}$ and $I_{C,2}$ (the critical current of $JJ_2$) are much higher than $I_{C,1}$. Figs. 2c,d show the measured $I_C(B)$ with different TG1. In such asymmetric SQUID ($I_{C,S} \gg I_{C,1}$), $I_C(B)$ reflects the current-phase relation (CPR) of the weaker

JJ$_1$. In this case, the center-to-peak amplitude of the $I_C(B)$ oscillation corresponds to $I_{C,1}$, while the center value corresponds to $I_{C,S}$[9]. When JJ$_1$ is almost pinched off ($I_{C,1}$ is small) and thus has negligible DOS, the gate voltage of TG1 only enhances $I_{C,1}$ (oscillation amplitude) but $I_{C,S}$ (oscillation center) remains the same, and the varying $V_{TG1}$ does not affect nonlocally JJ$_S$ (Fig. 2c, gray region in Fig. 2e). This is just the conventional asymmetric SQUID behaviors with independent junctions[9]. However, when the DOS of JJ$_1$, as well as $I_{C,1}$, is further increased by $V_{TG1}$, $I_{C,S}$ (oscillation center) rises significantly even with the fixed $V_{SG}$ (Fig. 2d, yellow region in Fig. 2e). Surprisingly, such nonlocal enhancement of $I_{C,S}$ by $V_{TG1}$ can be even larger than $I_{C,1}$ itself (Note that with such enhancement, the device is still in asymmetric SQUID regime, thus the amplitude and center of the $I_C(B)$ oscillation remain equal to $I_{C,1}$ and $I_{C,S}$, respectively). The extracted $I_{C,1}$ and $I_{C,S}$ versus $V_{TG1}$ is summarized in Fig. 2e, which clearly shows a regime of no nonlocal control of $I_{C,S}$ at lower $I_{C,1}$ ("non-molecule" regime, gray background) and another regime with significant nonlocal control of $I_{C,S}$ at higher $I_{C,1}$ ("molecule" regime, yellow background). We eliminate the possibility of the crosstalk between TG1, TG2, PG and SG by the following reasons: 1. the charge stability diagram of $I_C$ shows negligible dependence of SG on TG1, TG2 and PG (see Supplementary Information Sec. 2); 2. TG1, TG2 and PG are located far away from JJ$_S$, and in particular separated from JJ$_S$ by a grounded superconducting lead which has screening effect of the electric field from TG1, TG2 and PG; 3. the enhancement of $I_{C,S}$ by the nonlocal effect can be even larger than $I_{C,1}$ from the local gating. Such strongly nonlocal control of JJ$_S$ via gating the DOS of JJ$_1$ indicates the formation of the Andreev molecule between JJ$_1$ and JJ$_S$.

To further confirm the DOS-controlled Andreev molecule effect, we perform the tight-binding simulation[19] on a similar system, shown in Fig. 3a. To reduce the computational complexity, the nanowire is modeled as a 1d lattice with the middle section and S/D contacts as superconductor with finite Δ (drawn as solid circles) while the two normal sections without Δ represent JJ$_1$ and JJ$_S$ in Fig. 2 (drawn as empty circles), respectively. Similar to the experiment in Fig. 2, the length of the middle superconductor is set to be small and the Fermi level of JJ$_1$ ($U_{TG1}$) is stepped up in a range of values, acting as TG1. Meanwhile, the Fermi level of JJ$_S$ ($U_{SG}$) is fixed at a much higher level so that $I_{C,1} < I_{C,S}$ required for asymmetric SQUID. The phase drop across JJ$_1$ and JJ$_S$ are $\varphi_1$ and $\varphi_S$ respectively, and the CPR of both JJ$_1$ ($I_{JJ1}(\varphi_1)$) and JJ$_S$ ($I_{JJS}(\varphi_S)$) can be calculated at each $U_{TG1}$ via their respective Andreev levels (see Supplementary Information Sec. 3 for more details). The critical current $I_{C,1}$ and $I_{C,S}$ thus are the maximum values of the respective CPRs, and the SQUID critical current oscillation can be approximated by $I_{JJ1}(\varphi_1)+I_{C,S}$.

The simulation results of the SQUID oscillation with a series of $U_{TG1}$ are shown in Figs. 3b, c, with the extracted $I_{C,1}$ and $I_{C,S}$ in Fig. 3d. For more negative $U_{TG1}$, the nonlocal modulation of $I_{C,S}$ by $U_{TG1}$ is negligible (Figs. 3b, the gray region in Fig. 3d), similar to the non-molecule regime in Fig. 2c, e. However, when $U_{TG1}$ increases, $I_{C,S}$ is clearly modulated by $U_{TG1}$ (Figs. 3c, yellow region in Fig. 3d), qualitatively reproducing the molecule regime in Figs. 2d, e. By comparison between the measurement results in Fig. 2 and the numerical simulations in Fig. 3, we thus prove beyond doubt the existence of the new type Andreev molecule, that is, type II: the DOS (gate)-controlled Andreev molecule.

To have a better understanding of the physical mechanisms of such type II Andreev molecule, we show the typical Andreev levels of both JJ$_1$ and JJ$_S$ with $U_{TG1}$ = -3.5 (Figs. 3e,f, left panel, non-molecule regime) and $U_{TG1}$ = -0.7 (Figs. 3e,f, right panel, molecule regime), with the intermediate cases in Supplementary Information Fig. 3. For JJ$_1$ (Fig. 3e), the effect of $U_{TG1}$ is simply to locally tune its DOS: initially the ABS (that is, the levels with the energy lower than the induced gap around $\Delta$) is absent at $U_{TG1}$ = -3.5 (left panel), indicating negligible DOS of $JJ_1$, while the ABS appears by increasing $U_{TG1}$ to -0.7 (right panel) with the finite DOS of $JJ_1$[41].

For the nonlocal effects of $U_{TG1}$ on ABS of $JJ_S$ which is the focus here, there are two mechanisms, that is, the wavefunction hybridization and wavefunction transmission. First, the wavefunction hybridization between $JJ_1$ and $JJ_S$ conventionally manifests as the shift in $\varphi_S$ of the ABS of $JJ_S$ by $\varphi_1$ (a typical example is illustrated in Fig. 1b)[11-13,15,16,19]. By plotting the ABS of $JJ_S$ for $\varphi_1 = 0$ (solid lines in Fig. 3f) and for $\varphi_1 = \pi/4$ (dashed lines in Fig. 3f), the ABS of $JJ_S$ is almost unchanged by $\varphi_1$ for $U_{TG1} = -3.5$ (left panel in Fig. 3f), indicating that the wavefunction hybridization is negligible due to the vanishing DOS of $JJ_1$ shown in Fig. 3e (left panel). Meanwhile, the ABS of $JJ_S$ is significantly shifted by $\varphi_1$ for $U_{TG1} = -0.7$ (right panel in Fig. 3f). Such anomalous phase shift is typically seen in the previous Andreev molecules[11-13,15,16,19] indicates strong wavefunction hybridization when the ABS of $JJ_1$ appears with high DOS shown in Fig. 3e (right panel). Therefore, by increasing $U_{TG1}$, the wavefunction hybridization between $JJ_1$ and $JJ_S$ enhances, as illustrated in Fig. 3g.

Second, the wavefunction transmission of JJ is known to depend inversely on the size of the gap at $\varphi_S = \pi$ (marked by the double arrows in Fig. 3f)[41]. For $U_{TG1} = -3.5$ (Fig. 3f, left panel), the gap of the ABS of $JJ_S$ at $\varphi_S = \pi$ is large since the transmission of $JJ_S$ wavefunction into $JJ_1$ is suppressed to almost zero due to depleted DOS (Fig. 3h, left panel), whereas such transmission becomes higher and tunable with finite DOS in $JJ_1$ at $U_{TG1} = -0.7$ (Fig. 3h, right panel). Therefore, $U_{TG1}$ also nonlocally tune the the wavefunction transmission of $JJ_S$ into $JJ_1$ as illustrated in Fig. 3h.

In short, these two non-local gate-tunable mechanisms (wavefunction hybridization and transmission) both contribute to the nonlocal control of $I_{C,S}$ by $U_{TG1}$. We note that the simulations in Fig. 3b-d of JJ$_S$ (or JJ$_1$) are done with $\varphi_1 = 0$ (or $\varphi_S = 0$), while the qualitatively similar behaviors are reproduced also for other finite values of $\varphi_{1,S}$ (see Supplementary Information Fig. 4). This highlights the validity of the type II Andreev molecule without the requirement of specific phase control.

The advantage of the type II DOS (gate)-controlled Andreev molecule compared to the conventional type I phase-controlled Andreev molecules[11-13,15,16,19] lies particularly in the fact that it uses the local gate instead of the superconducting loop for each site for controlling and thus has more compact design and better extensibility. Moreover, it is also worth noting that the DOS (gate)-controlled Andreev molecule shares close similarity as the two-site Kitaev chain[28]. In particular, the above gate-dependent molecule effects are highly relevant for the fine tuning of the local finger gates which carefully adjusts the inter-site wavefunction couplings to create the poorman's MZMs[28]. Our results thus highlight the importance of the DOS (gate)-controlled Andreev molecule effects in the Kitaev chains, which should be explicitly considered for future experimental and theoretical investigations.

**Extended Andreev molecule**

In the following, we shall demonstrate the nonlocal DOS-controllability in Andreev molecule with several more complex configurations by tuning TG1, TG2 and PG in one site (left site), while the other site (right site) remains as JJ$_S$ with stronger critical current. First, as a natural extension of the simple molecule with two sites, the left site can be configured as two junctions connecting in series, thus creating a three-sites molecule between JJ$_1$, JJ$_2$, JJ$_S$.

Here, TG1 and TG2 are such that $I_{C,1}$ is larger than $I_{C,2}$. $I_{C,S}$ is still much larger than $I_{C,2}$ as in Fig. 2, enabling the compact readout via asymmetric SQUID. We note that different from Fig. 2 now $JJ_2$ instead $JJ_1$ is the weakest JJ (Fig. 4a).

When turning up $JJ_2$ via TG2, we observe the similar trends in Figs. 4c-e as Figs. 2c-e: When the $I_{C,2}$ is small (Fig. 4c, gray region in Fig.4e) $I_{C,S}$ is independent of TG2, while $I_{C,S}$ is significantly modulated nonlocally by TG2 when $I_{C,2}$ is sufficiently large (Fig. 4d, yellow region in Fig.4e). Such nonlocal control of $I_{C,S}$ via TG2 several micrometers away can be explained in the two possibilities illustrated in Fig. 4b:

(A) $JJ_2$ and $JJ_1$ are also coupled by the Andreev molecule effects and thus the three junctions ($JJ_2$, $JJ_1$, $JJ_S$) form an extended Andreev molecule, as illustrated in the upper panel of Fig. 4b. In such case, TG2 nonlocally controls the ABSs of $JJ_1$ which causes further nonlocal control of ABSs of $JJ_S$. In three-sites Kitaev chain, the superconducting Al films between $JJ_2$ and $JJ_1$ and between $JJ_1$ and $JJ_S$ should be grounded[28-30], whereas here the Al film between $JJ_2$ and $JJ_1$ is not grounded. However, since the grounding does not hinder the wavefunction hybridization and transmission between the adjacent JJs, this setup is similar to three-site Kitaev chain configuration and the DOS (gate)-controlled Andreev molecule effects should also be taken into consideration in the multi-sites Kitaev chains.

(B) $JJ_2$ and $JJ_1$ are independent junctions connected in series, as illustrated in the bottom panel of Fig. 4b. Since $I_{C,2} < I_{C,1}$, the supercurrent through $JJ_1$ is limited by the smaller $I_{C,2}$. Therefore, when $JJ_2$ reaches its own $I_{C,2}$, $JJ_1$ is effectively phase-biased with $\varphi_1 = I_{JJ1}^{-1}(I_{C,2})$, where $I_{JJ1}^{-1}$ is the inverse CPR of the $JJ_1$. Since TG2 controls $I_{C,2}$, it indirectly modifies $\varphi_1$, which consequently affects $I_{C,S}$ via the Andreev molecule effect between $JJ_1$ and $JJ_S$.

We note that both (A) and (B) can be used to construct the extended Andreev molecule. Meanwhile, in Supplementary Information Sec. 4, we discuss the higher likelihood of (A), thus the Andreev molecule coupling between three sites $JJ_2$, $JJ_1$, $JJ_S$, inferred from the anomalous SQUID oscillations of $JJ_2$.

**Andreev molecule sensor for single Cooper pair states**

In the second case, we configure the left site of the molecule as the Cooper pair transistor (CPT) while still keeping the right site as $JJ_S$, and will show how to use the Andreev molecule between $JJ_S$ and CPT as a non-local (thus non-invasive) and sensitive sensor of the Cooper pair states of the CPT. The CPT is formed as follows. In the left site, the section of InAs nanowire covered with epitaxial Al (that is, "island") linking $JJ_1$ and $JJ_2$ may have significant charging energy $E_C$ due to its small size and the presence of the two barriers in the end (Fig. 5a). When the device has significant $E_C$ by tuning TG1 and TG2, single Cooper pairs can tunnel sequentially through the island due to Coulomb blockade effect, forming a Cooper-pair transistor (CPT)[43]. The CPT is a well-understood quantum device widely used in constructing transmon qubit[37,38,44] and detecting Majorana zero modes in hybrid systems[45,46]. The conductance of the CPT is maximal if PG is such that the Fermi level of the island is aligned with those of the S and D ("on Coulomb resonance", red dashed lines in Fig. 5b), while it is minimal if the Fermi level of the island is misaligned ("Coulomb blockaded", green solid lines in Fig. 5b). With the molecule effect between the entire CPT and $JJ_S$, the ABSs of $JJ_S$ can be nonlocally modulated by the single Cooper pair occupation on the island which produces a nonlocal oscillating response of $I_{C,S}$ by $V_{PG}$. Therefore by monitoring such nonlocal response, the charge states of the CPT can be non-locally detected.

The strong Coulomb blockade effect from $E_C$ of the CPT is directly verified by the measurement of differential conductance $G = dI/dV$ as functions of $V_{PG}$ and the dc voltage bias $V_{DC}$, which exhibits the well-established features of "Coulomb diamond"[43] (marked by dashed lines) in Fig. 5c. This measurement is performed with an in-plane field $B = 0.2$ T, which is applied to destroy the superconductivity of the thicker Al contacts but still keeps the superconductivity of the thin Al of the island, and thus eliminates the influence of the Andreev molecule effects and the parallel superconducting channel from JJ$_S$. The charge number of the superconducting island is independently controlled by PG. At $V_{DC} = 0$ inside the superconducting gap (bottom panel), $G(V_{PG})$ oscillates with $|2e|$ periodicity, is high on Coulomb resonance, and is low at Coulomb blockade[46]. Meanwhile, when $V_{DC}$ is larger than the gap (middle panel), the oscillation periodic of $G(V_{PG})$ is halved, reflecting the $|1e|$ transport due to quasiparticles[46]. By comparing the oscillation periodicity between high and zero bias voltages, the left site is indeed a CPT with single Cooper pair (that is $|2e|$ period ≈ 0.03 V) transport. As a standard procedure[43,46], the charging energy $E_C = e^2/2C_\Sigma \approx 21$ μeV is estimated by the total height of the diamond which is $16E_C$ in the $|2e|$-oscillation regime.

Removing the added field with the similar gate settings and again measuring the SQUID critical current, Fig. 5d shows the typical $I_C(B)$ oscillation with out-of-plane flux between on and off the Coulomb resonance. Similar to previous configurations, the SQUID is asymmetric, and the center value of the oscillation thus reflects $I_{C,S}$ of JJ$_S$. Again, clear shift of center value (that is, $I_{C,S}$) is seen between resonance (red dot) and blockaded (green dot) cases, by the nonlocal gating of PG. Fig. 5e shows the extracted $I_{C,S}$ (also from the center value of $I_C(B)$) versus $V_{PG}$, showing maximum (minimum) value on resonance (off resonance), in agreement with the CPT behaviors[37,38,44]. It is surprising that even a small DOS change associated with single Cooper pair addition/extraction to the island can sensitively and nonlocally modulate a JJ with a much larger critical current. Indeed, if we define the gate sensitivity of the molecule as $S = \Delta I_{C,S}/\Delta V_{TG,PG}$, we find that $S \approx 18$nA/1V = 18 nA/V in Fig. 2e (estimated between $V_{TG1} \approx 3$V and 4V) while $S \approx 5.8$ nA/0.016 V = 362 nA/V in the CPT here in Fig. 5, with an over 20 times enhancement. This thus demonstrates the high sensitivity of the DOS-controlled Andreev molecule, which may be used for constructing a single-electron/Cooper pair charge sensor[10,32,33]. We prove here, besides charge sensing, the interferometric readout technique with a SQUID loop can also be useful for sensing the parity of MZMs in potential topological qubit applications[34,35].

**Conclusions**

In conclusion, we have demonstrated a new way of controlling the Andreev molecule not by the conventional type I phase scheme with multiple superconducting loops but by the DOS via the compact design of electrical gating (type II scheme). A single SQUID loop for the whole molecule is employed for simultaneous readout of the critical current for multiple sites. Based on the Al-InAs nanowire hybrid SQUID device, we find that the Andreev molecule manifests as the strong nonlocal enhancement of critical current of one molecule site by the gate-controlled DOS of the other site, which has been qualitatively reproduced by tight-binding simulations. We find both wavefunction hybridization and transmissions cause the non-local response of $JJ_S$ supercurrent by the $JJ_1$ gate. With the gate configurations, we further develop such DOS-controlled scheme to more complex device structures such as an extended Andreev molecule chain with multiple sites and an Andreev molecule sensor with an over 20 times enhancement in sensitivity and able to nonlocally sense a single addition/extraction of the Cooper pair. The interferometric readout technique also has potential applications for parity sensing of the Cooper pairs as well as single particles, useful for topological quantum computing[34,35].

In addition, how the nonlocal effects respond to the gate between different sites coupled through hybrid superconducting contacts should be explicitly considered in the construction of the recently demonstrated Kitaev chain and the fine-tuning of the sweet point for the poorman's MZMs[28,29], in particular when such Kitaev chain further extends into long-chain structure with multiple sites[30] via complex tuning knobs to realize topological protection. Moreover, the combination of compact design without multiple superconducting loops, straightforward extensibility to complex devices, and high sensitivity to the charge/parity of a nonlocal site make such novel DOS (gate)-controlled Andreev molecule not only highly relevant for engineering topological ABSs via multijunction devices[21-26], but also greatly suitable for non-invasive charge/parity sensing as well as sensitive quantum signal transmuting in the future construction and readout of large-scale Majorana or Kitaev-chain based topological qubit systems[27-30,32,33].

**Method**

*InAs-Al nanowire growth*: InAs nanowires were epitaxially grown on commercial n-type Si(111) substrates using a solid-source molecular-beam epitaxy (MBE) system (VG V80H), with Ag nanoparticles employed as the catalysts. Prior to MBE loading, the Si substrates were subjected to a chemical pretreatment step involving immersion in a 2% diluted hydrofluoric acid solution for 1 minute to eliminate surface contaminants and native oxide layers, as reported elsewhere[47]. For catalyst deposition, an ultrathin Ag film with a nominal thickness of <0.5 nm was thermally evaporated onto the Si(111) substrates at room temperature within the MBE growth chamber. Subsequent *in situ* annealing at 550 °C for 20 minutes induced dewetting of the Ag film, resulting in the formation of discrete Ag nanoparticles with dimensions suitable for nanowire nucleation. InAs nanowires were then grown for 80 minutes at a substrate temperature of 485 °C, utilizing an arsenic-to-indium beam equivalent pressure (BEP) ratio of ~42 (corresponding to BEP fluxes of $1.1 \times 10^{-7}$ mbar for In and $4.6 \times 10^{-6}$ mbar for $As_4$). This process yielded ultrathin InAs nanowires with diameters ranging from ~20 to 40 nm. Following InAs growth, the sample was transferred to a preparation chamber at 300 °C to prevent arsenic condensation on the nanowire surfaces. The substrate was subsequently cooled to a low temperature (~−40 °C) via a combination of natural cooling and liquid nitrogen-assisted cooling, as detailed in prior reports[48]. Al deposition was performed by evaporating from a Knudsen cell at an oblique angle of ~20° relative to the substrate normal (~70° from the substrate surface) and a cell temperature of ~1150 °C for 100 seconds, yielding a deposition rate of approximately 0.08 nm/s. To achieve conformal half-shell Al coatings, substrate rotation was disabled during Al growth, ensuring unidirectional deposition. Upon completion of the InAs-Al heterostructure synthesis, the sample was rapidly extracted from the MBE chamber and exposed to ambient conditions for natural oxidation, stabilizing the Al shell. This process enables the fabrication of InAs-Al core-shell nanowires with controlled dimensions and interface properties, suitable for applications in quantum electronic devices.

*Device fabrication*: InAs-Al nanowires were transferred by a wiper. The wiper first gently swiped the growth substrate and then swiped again on the device substrate which was highly p-doped Si covered by 300 nm silicon dioxide. The randomly deposited nanowires are then selected via scanning electron microscopy (SEM) with minimal exposure time. With standard electron beam lithography processes, the contact areas were defined. Tunnel barriers in the nanowire are formed by etching the aluminum (Al) thin film using Transene Aluminum Etchant Type D at 50°C for 10 seconds. Ohmic contacts to the InAs nanowire were fabricated by 80 s Ar plasma etching at a power of 50 W and pressure of 0.05 Torr, followed by metal deposition of Ti/Al (5/65 nm) bilayer. We note the same device used in this work has been measured in a separate work using the different dataset by the same authors[48].

*Measurement technique*: The fast counter measurement technique are explained in details in Supplementary Information Sec. 1.

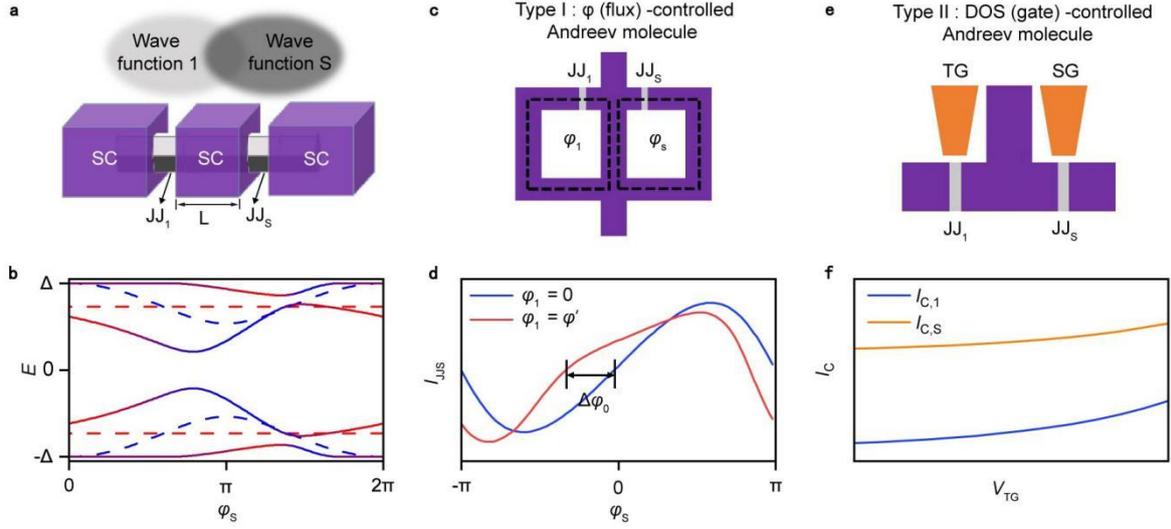

**Fig. 1. "Type I" and "Type II" Andreev molecules**: **a,** Schematic of the two spatially overlapped wavefunctions of the ABSs from junctions $JJ_1$ and $JJ_S$ forming an Andreev molecule. **b,** Example of the independent ABSs of $JJ_1$ (red dashed lines, set by $\varphi_1$ and independent of $\varphi_S$) and of $JJ_S$ (blue dashed lines, depending on $\varphi_S$) and the hybridized ABSs (solid lines with the hybridization happening at the level crossings). Time-reversal symmetry may be broken due to hybridization. **c, d,** Type I: phase-controlled Andreev molecule with two superconducting rings tuning $\varphi_{1,S}$ of $JJ_{1,S}$. The CPR of $JJ_S$ in **d** is modulated nonlocally by $\varphi_1$. An anomalous phase shift $\Delta\varphi_0$ is introduced for finite $\varphi_1$. **e, f,** Type II: DOS-controlled Andreev molecule with two gates TG, SG tuning the ABS density of $JJ_{1,S}$. Different from **c**, the SQUID loop is not required to control the molecule. The molecule effect manifests as the nonlocal tuning of $I_{C,S}$ in $JJ_S$ by TG of $JJ_1$ in **f**.

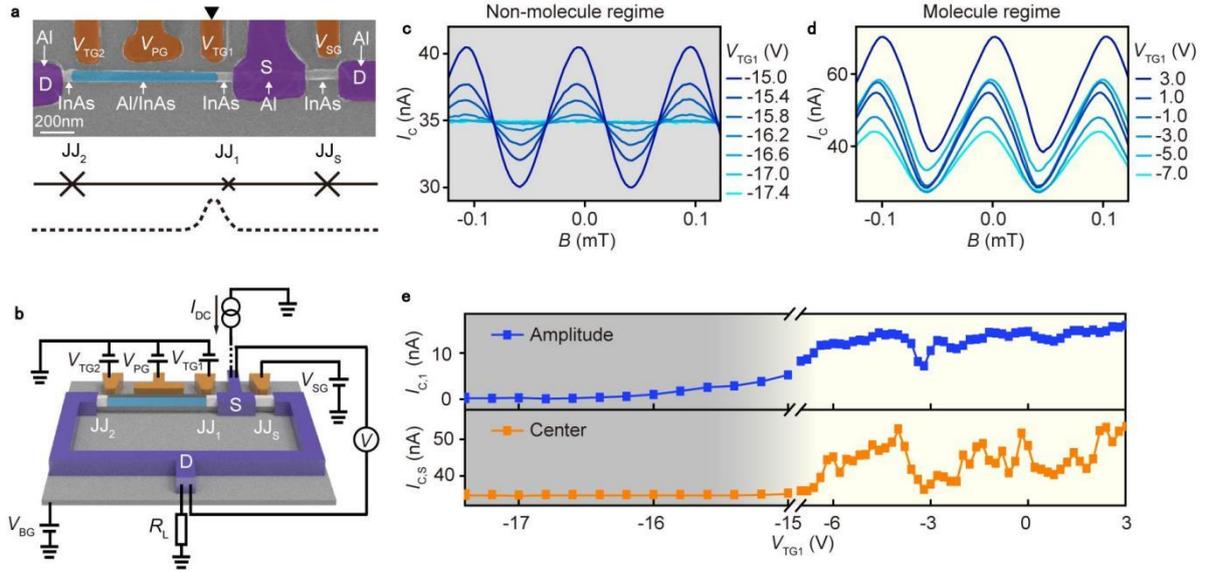

**Fig. 2. Nonlocal critical current response in DOS (gate)-controlled molecule**: **a,** False-color SEM image of the Andreev molecule device. Scale bar: 200 nm. The epitaxial Al (light blue) is removed to expose bare InAs (light gray) as $JJ_1$, $JJ_2$ and $JJ_S$. The length of the middle Al section (purple, marked as S) is ~ 300 nm. **b,** Schematic of the measurement setup. **c, d,** Measured $I_C(B)$ for different $V_{TG1}$ (marked by black triangle in **a**). Global backgate $V_{BG} = 0$ V; $V_{TG2} = 6$ V, $V_{SG} = 0$ V, such that $I_{C,1} \ll I_{C,2}, I_{C,S}$. **c:** Non-molecule regime (gray background): For more negative $V_{TG1}$, the oscillation amplitude ($I_{C,1}$) increases while its center value ($I_{C,S}$) is unchanged. $JJ_1$ and $JJ_S$ are independent junctions. **d:** Molecule regime (yellow background): For negative to positive $V_{TG1}$, the oscillation amplitude ($I_{C,1}$) increases while its center value ($I_{C,S}$) also increases. The nonlocal response of $I_{C,S}$ to TG1 indicates the Andreev molecule effects between $JJ_1$ and $JJ_S$. **e,** Extracted center-to-peak amplitude ($I_{C,1}$) and center value ($I_{C,S}$) from the measured $I_C(B)$ versus $V_{TG1}$ showing the transition from the non-molecule (gray background) to the molecule regime (yellow background).

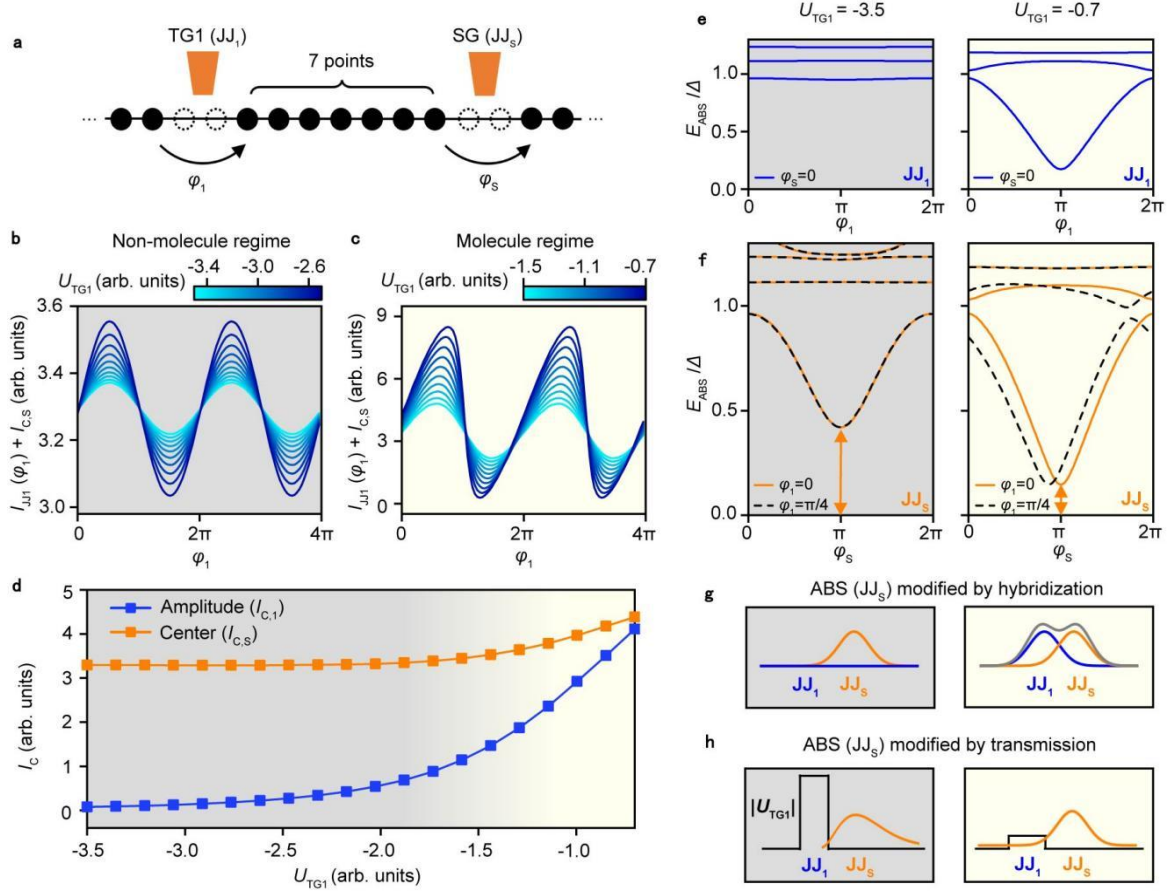

**Fig. 3. Tight-binding calculation of the DOS (gate)-controlled Andreev molecule: a,** The 1d lattice for Andreev molecule. Solid (hollow) dots: superconducting (normal) section. The middle superconducting section has 7 lattices, shorter than the coherence length. The Fermi level set by $U_{TG1}$ is stepped while $U_{SG}$ is fixed such that $I_{C,1} < I_{C,S}$, similar to Fig. 2. **b, c** Simulated $I_{JJ1}(\varphi_1)+I_{C,S}$ with different $U_{TG1}$ for non-molecule (**b**) and molecule regimes (**c**), similar to Figs. 2c,d. $\varphi_1$ across $JJ_1$ grows proportionally with the SQUID flux $\Phi$. $I_{JJ1}(\varphi_1)$ is calculated with $\varphi_S = 0$; $I_{C,S}$ is calculated with $\varphi_1 = 0$. **d,** $I_{C,1}$ and $I_{C,S}$ versus $U_{TG1}$ showing also showing clear nonlocal response of $I_{C,S}$ to $U_{TG1}$, similar to Figs. 2e. The density of points in $U_{TG1}$ is less than **b,c** for clarity. **e,** Calculated Andreev levels of $JJ_1$ versus $\varphi_1$ with $\varphi_S = 0$ (blue solid lines). Only positive and low energy is shown for clarity. Left panel: $U_{TG1} = -3.5$, The ABS in $JJ_1$ is negligible due to vanishing DOS. Right panel: $U_{TG1} = -0.7$, ABS appears in $JJ_1$

with higher DOS. **f,** Calculated Andreev levels of $JJ_S$ versus $\varphi_S$ with $\varphi_1 = 0$ (orange solid lines) and with $\varphi_1 = \pi/4$ (dashed lines). Left panel: $U_{TG1} = -3.5$, negligible phase shift between solid and dashed lines indicating vanishing hybridization between $JJ_1$ and $JJ_S$ (see **g,** left panel). The larger gap at $\varphi_S = \pi$ (marked by double arrow) indicates suppressed wavefunction transmission of $JJ_S$ into $JJ_1$ (see **f,** left panel). Right panel: $U_{TG1} = -0.7$, significant shift between solid and dashed lines indicating strong hybridization between $JJ_1$ and $JJ_S$ (see **g,** right panel). The smaller gap at $\varphi_S = \pi$ (marked by double arrow) indicates high wavefunction transmission of $JJ_S$ into $JJ_1$ (see **f,** right panel). **g,** illustrations of ABS wavefunctions modified by hybridization. Left panel: vanishing hybridization with $U_{TG1} = -3.5$. Right panel: strong hybridization with $U_{TG1} = -0.7$. Blue (yellow) lines: wavefunctions of $JJ_1$ ($JJ_S$) before hybridization. Gray line: wavefunction after hybridization. **h,** illustrations of ABS wavefunctions modified by transmission. Left panel: suppressed transmission into $JJ_1$ with $U_{TG1} = -3.5$. Right panel: high transmission into $JJ_1$ enabled with $U_{TG1} = -0.7$.

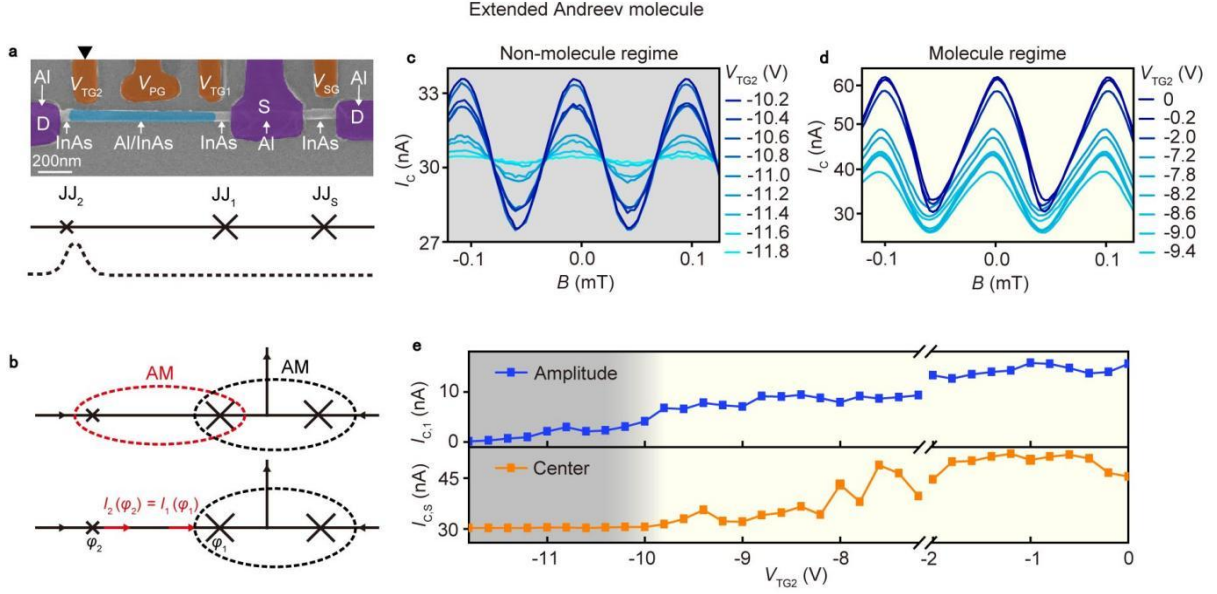

**Fig. 4. Nonlocal critical current response by extended Andreev molecule: a,** Same as Fig. 2a but with fixed $V_{TG1}$ and varying $V_{TG2}$ (marked by black triangle). $I_{C,2} \ll I_{C,1}, I_{C,S}$. **b,** Upper: $JJ_2$ forms an additional Andreev molecule with $JJ_1$, which is already in the molecule with $JJ_S$. The device thus realizes an Andreev molecule chain. Lower: No Andreev molecule between $JJ_1$ and $JJ_2$. The nonlocal control of $I_{C,S}$ by $V_{TG2}$ is via supercurrent in series $I_{JJ1}(\varphi_1) = I_{JJ2}(\varphi_2)$ (see text). **c, d,** Measured $I_C(B)$ for different $V_{TG2}$. $V_{BG} = 0$ V; $V_{TG1} = 8$ V, $V_{SG} = 0$ V. Similar to Fig. 2c, d, for less negative $V_{TG2}$, the oscillation amplitude ($I_{C,2}$) increases while its center value ($I_{C,S}$) also significantly increases. **e,** Extracted $I_{C,2}$ and $I_{C,S}$ versus $V_{TG2}$ from the measured $I_C(B)$.

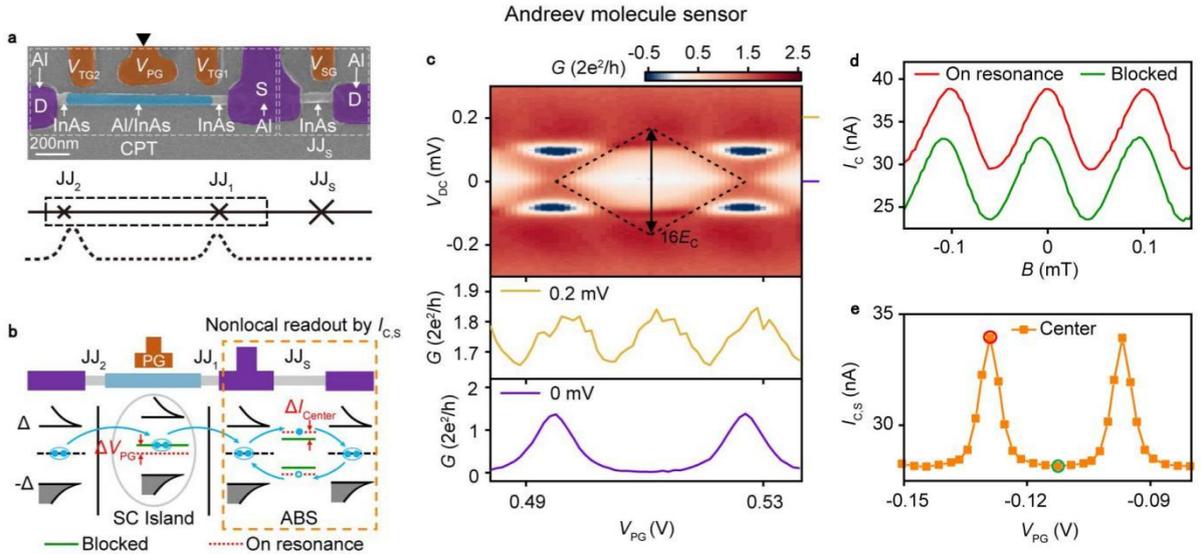

**Fig. 5. Nonlocal critical current response in Andreev molecule sensor: a,** Same as Fig. 2a but with fixed $V_{TG1,2}$ such $E_C$ is significant. $JJ_{1,2}$ and the superconducting island in between (epitaxial Al on InAs nanowire, light blue) thus form a CPT. Varying $V_{PG}$ (marked by black triangle) controls the charge number of the island. $V_{BG}$ = -9.00 V; $V_{TG1}$ = 0.22 V, $V_{TG2}$ = 0 V, $V_{SG}$ = 0 V. $I_{C,CPT} \ll I_{C,S}$. **b,** The CPT state (on Coulomb resonance and Coulomb blockaded, dashed/solid lines respectively) modulates the ABS of $JJ_S$ (dashed/solid lines respectively), via the molecule effects. **c,** Top panel: $G(V_{PG}, V_{DC})$, showing Coulomb diamonds (dashed lines) and its height as $16E_C$. $E_C \approx 21$ μeV. Middle and bottom panels: $G(V_{PG})$ with $V_{DC}$ =0 (|2e| periodicity) and $V_{DC}$ = 300 μV (|1e| periodicity). In-plane field $B$ = 0.2 T. **d,** Measured $I_C(B)$ near $B$ = 0 T with on resonance (red) and in blockade (green), respectively. $V_{BG}$ = -5.09 V; $V_{TG1}$ = 0.15 V, $V_{TG2}$ = -2.35 V, $V_{SG}$ = 0 V. **e,** Extracted $I_{C,S}(B)$ for each $V_{PG}$, showing large nonlocal gate sensitivity $S = \Delta I_{C,S}/\Delta V_{PG} \approx 5.8$ nA/0.016 V = 362 nA/V.

**Data availability**

All data needed to evaluate the conclusions in the paper are present in the main text and/or the supplementary information. Raw data generated in this study are available from the corresponding author upon reasonable request.

**Acknowledgement**

We are grateful to Chunxiao Liu for helpful discussions. The work of Z.D. and J.S. was supported by the Young Scientists Fund of the National Natural Science Foundation of China (Grant No. 2024YFA1613200) The work of D.P. and J.Z. was supported by the National Natural Science Foundation of China (Grant Nos. 12374459, 61974138 and 92065106), the Innovation Program for Quantum Science and Technology (Grant 2021ZD0302400). D. P. acknowledges the support from Youth Innovation Promotion Association, Chinese Academy of Sciences (Nos. 2017156 and Y2021043). The work of J.S., L.L., F.Q. and G.L. were supported by the National Key Research and Development Program of China (Grant Nos. 2023YFA1607400), the Beijing Natural Science Foundation (Grant No. JQ23022), the Strategic Priority Research Program B of Chinese Academy of Sciences (Grant No. XDB33000000), the National Natural Science Foundation of China (Grant Nos. 12174430, 92365302), and the Synergetic Extreme Condition User Facility (SECUF, https://cstr.cn/31123.02.SECUF). Y.L. acknowledges support from National Natural Science Foundation of China, Grant No. 12404154. The work of other authors were supported by the National Key Research and Development Program of China (Grant Nos. 2019YFA0308000, 2022YFA1403800, 2023YFA1406500, and 2024YFA1408400), the

National Natural Science Foundation of China (Grant Nos. 12274436, 12274459), the Beijing Natural Science Foundation (Grant No. Z200005), and the Synergetic Extreme Condition User Facility (SECUF, https://cstr.cn/31123.02.SECUF). The work is also funded by Chinese Academy of Sciences President's International Fellowship Initiative (Grant No. 2024PG0003).

**Author contributions**

J.S. and Z.D. conceived and designed the experiment

S.F., G.L. fabricated the devices and performed the transport measurements, with assistance from Y.L., A.W., Z.Z., X.G., X.D., discussed with B.T., Z.L., P.L., F.Q., G.L, supervised by Z.D., L.L., J.S.

D.P. and J.Z. grew nanowire materials

Z.D. and Y.S. performed the simulations and data analysis, discussed with J.S., S.F., G.L.

Y. S. and S.F. prepared the figures, supervised by Z.D., J.S.

Z.D. and J.S. wrote the manuscript, with input from all authors.

**Competing interests**

Authors declare that they have no competing interests.

**References**

1. C. Nayak, S. H. Simon, A. Stern, M. Freedman, and S. Das Sarma, Non-Abelian anyons and topological quantum computation, Rev. Mod. Phys. 80, 1083 (2008).

2. D. Aasen, M. Hell, R. V. Mishmash, A. Higginbotham, J. Danon, M. Leijnse, T. S. Jespersen, J. A. Folk, C. M. Marcus, K. Flensberg, and J. Alicea, Milestones toward Majorana-based quantum computing, Phys. Rev. X 6, 031016 (2016).

3. B. Lian, X. Sun, A. Vaezi, X. Qi, & S. Zhang, Topological quantum computation based on chiral Majorana fermions, Proc. Natl. Acad. Sci. 115, 43 (2018).

4. Liang Fu and C. L. Kane, Superconducting proximity effect and Majorana fermions at the surface of a topological insulator, Phys. Rev. Lett. 100, 096407 (2008).

5. J. Li, T. Neupert, B. A. Bernevig, and A. Yazdani, Two-dimensional chiral topological superconductivity in Shiba lattices. Nat Commun 7, 12297 (2016).

6. S. Nadj-Perge, I. K. Drozdov, J. Li, H. Chen, S. Jeon, J. Seo, A. H. MacDonald, B. A. Bernevig, and A. Yazdani, Observation of Majorana fermions in ferromagnetic atomic chains on a superconductor. Science, 346, 602-607 (2014).

7. Flensberg, K., von Oppen, F. & Stern, A. Engineered platforms for topological superconductivity and Majorana zero modes. Nat Rev Mater 6, 944–958 (2021).

8. R. M. Lutchyn, E. P. A. M. Bakkers, L. P. Kouwenhoven, P. Krogstrup, C. M. Marcus and Y. Oreg. Majorana zero modes in superconductor–semiconductor heterostructures. Nat Rev Mater 3, 52–68 (2018).

9. M. Tinkham: Introduction to Superconductivity. International series in pure and applied physics. McGraw Hill, New York (1996).

10. Z. Su, A. B. Tacla, M. Hocevar, D. Car, S. R. Plissard, E. P. A. M. Bakkers, A. J. Daley, D. Pekker, and S. M. Frolov Andreev molecules in semiconductor nanowire double quantum dots Nat. Commun. 8, 585 (2017).

11. D. Z. Haxell, M. Coraiola, M. Hinderling, S. C. ten Kate, D. Sabonis, A. E. Svetogorov, W. Belzig, E. Cheah, F. Krizek, R. Schott, W. Wegscheider, and F. Nichele, Demonstration of the nonlocal Josephson effect in Andreev molecules, Nano Letters 23,16, 7532-7538 (2023).

12. J.-D. Pillet, V. Benzoni, J Griesmar, . J.-L. Smirr, Ç. O. Girit, Nonlocal Josephson effect in Andreev molecules. Nano Lett. 19, 7138– 7143 (2019).

13. V. Kornich, H. S. Barakov, and Y. V. Nazarov, Fine energy splitting of overlapping Andreev bound states in multiterminal superconducting nanostructures. Phys. Rev. Res. 1, 033004 (2019).

14. O. Kürtössy, Z. Scherübl, G. Fülöp, I. E. Lukács, T. Kanne, J. Nygård, P. Makk, S. Csonka, Andreev molecule in parallel InAs nanowires. Nano Lett. 21, 7929– 7937 (2021).

15. S. Matsuo, T. Imoto, T. Yokoyama, Y. Sato, T. Lindemann, S. Gronin, G. C. Gardner, S. Nakosai, Y. Tanaka, M. J. Manfra and S. Tarucha, Phase-dependent Andreev molecules and superconducting gap closing in coherently-coupled Josephson junctions, Nat. Commun. 14, 8271 (2023).

16. S. Matsuo, T. Imoto, T. Yokoyama, Y. Sato, T. Lindemann, S. Gronin, G. C. Gardner, M. J. Manfra, and S. Tarucha. Josephson diode effect derived from short-range coherent coupling. Nat. Phys. 19, 1636–1641 (2023).

17. J.-D. Pillet, S. Annabi, A. Peugeot, H. Riechert, E. Arrighi, J. Griesmar, and L. Bretheau, Josephson diode effect in Andreev molecules, Phys. Rev. Research 5, 033199 (2023).

18. S. Matsuo, R. S. Deacon, S. Kobayashi, Y. Sato, T. Yokoyama, T. Lindemann, S. Gronin, G. C. Gardner, K. Ishibashi, M. J. Manfra and S. Tarucha, Shapiro response of superconducting diode effect derived from Andreev molecules, Phys. Rev. B 111, 094512 (2025).

19. S. Matsuo, T. Imoto, T. Yokoyama, Y. Sato, T. Lindemann, S. Gronin, G. C. Gardner, M. J.

Manfra, S. Tarucha, Phase engineering of anomalous Josephson effect derived from Andreev molecules, Sci. Adv. 9, eadj3698 (2023).

20. M. Gupta, G. V. Graziano, M. Pendharkar, J. T. Dong, C. P. Dempsey, C. Palmstrøm, and V. S. Pribiag, Gate-tunable superconducting diode effect in a three-terminal Josephson device, Nat. Commun. 14, 3078 (2023)

21. M. Coraiola, D. Z. Haxell, D. Sabonis, H. Weisbrich, A. E. Svetogorov, M. Hinderling, S. C. ten Kate, E. Cheah, F. Krizek, R. Schott, W. Wegscheider, J. C. Cuevas, W. Belzig, and F. Nichele, Phase-engineering the Andreev band structure of a three-terminal Josephson junction, Nat. Commun. 14, 6784 (2023).

22. W. Jung, S. Jin, S. Park, S.-H. Shin, K. Watanabe, T. Taniguchi, G. Y. Cho, and G.-H. Lee, Tunneling spectroscopy of Andreev bands in multiterminal graphene-based Josephson junctions, Sci. Adv., 11, 21 (2025).

23. R.-P. Riwar, M. Houzet, J. S. Meyer, and Y. V. Nazarov, Multi-terminal Josephson junctions as topological matter, Nat. Commun. 7, 11167 (2016).

24. C. Padurariu, T. Jonckheere, J. Rech, R. Mélin, D. Feinberg, T. Martin, and Yu. V. Nazarov, Closing the proximity gap in a metallic Josephson junction between three superconductors Phys. Rev. B 92, 205409 (2015).

25. R. L. Klees, G. Rastelli, J. C. Cuevas, and W. Belzig, Microwave spectroscopy reveals the quantum geometric tensor of topological Josephson matter, Phys. Rev. Lett. 124, 197002 (2020).

26. J. S. Meyer and M. Houzet, Nontrivial Chern numbers in three-terminal Josephson junctions, Phys. Rev. Lett. 119, 136807 (2017).

27. G. Wang, T. Dvir, G. P. Mazur, C.-X. Liu, N. van Loo, S. L. D. ten Haaf, A. Bordin, S. Gazibegovic, G. Badawy, E. P. A. M. Bakkers, M. Wimmer and L. P. Kouwenhoven, Singlet and triplet Cooper pair splitting in hybrid superconducting nanowires, Nature, 612,

448–453 (2022).

28. T. Dvir, G. Wang, N. van Loo, C.-X. Liu, G. P. Mazur, A. Bordin, S. L. D. ten Haaf, J.-Y. Wang, D. van Driel, F. Zatelli, X. Li, F. K. Malinowski, S. Gazibegovic, G. Badawy, E. P. A. M. Bakkers, M. Wimmer and L. P. Kouwenhoven, Realization of a minimal Kitaev chain in coupled quantum dots, Nature 614, 445–450 (2023)

29. S. L. D. ten Haaf, Q. Wang, A. M. Bozkurt, C.-X. Liu, I. Kulesh, P. Kim, D. Xiao, C. Thomas, M. J. Manfra, T. Dvir, M. Wimmer and S. Goswami, A two-site Kitaev chain in a two-dimensional electron gas, Nature, 630, 329–334 (2024)

30. A. Bordin, C.-X. Liu, T. Dvir, F. Zatelli, S. L. D. ten Haaf, D. van Driel, G. Wang, N. van Loo, Y. Zhang, J. C. Wolff, T. Van Caekenberghe, G. Badawy, S. Gazibegovic, E. P. A. M. Bakkers, M. Wimmer, L. P. Kouwenhoven and G. P. Mazur, Enhanced Majorana stability in a three-site Kitaev chain, Nature Nanotechnology (2025).

31. M. Kocsis, Z. Scherübl, G. Fülöp, P. Makk, and S. Csonka, Strong nonlocal tuning of the current-phase relation of a quantum dot based Andreev molecule, Phys. Rev. B 109, 245133 (2024).

32. D. van Driel, B. Roovers, F. Zatelli, A. Bordin, G. Wang, N. van Loo, J. C. Wolff, G. P. Mazur, S. Gazibegovic, G. Badawy, E.P.A.M. Bakkers, L. P. Kouwenhoven, and T. Dvir, Charge sensing the parity of an Andreev molecule, PRX Quantum 5, 020301 (2024).

33. Z. Scherübl, A. Pályi and S. Csonka, Transport signatures of an Andreev molecule in a quantum dot–superconductor–quantum dot setup, Beilstein J. Nanotechnol., 10, 363–378 (2019).

34. Microsoft Azure Quantum., M. Aghaee, A. Alcaraz Ramirez, et al. Interferometric single-shot parity measurement in InAs–Al hybrid devices. Nature 638, 651–655 (2025).

35. J.-Y. Wang, C. Schrade, V. Levajac, D. van Driel, K. Li, S. Gazibegovic, G. Badawy, R. L. M. Op het Veld, J. S. Lee, M. Pendharkar, C. P. Dempsey, C. J. Palmstrøm, E. P. A. M.

Bakkers, L. Fu, L. P. Kouwenhoven, and J. Shen, Supercurrent parity meter in a nanowire Cooper pair transistor. Sci. Adv. 8,eabm9896 (2022).

36. G.P. Mazur, N. van Loo, D. van Driel, J.-Y. Wang, G. Badawy, S. Gazibegovic, E.P.A.M. Bakkers, L.P. Kouwenhoven, Gate-tunable Josephson diode, Phys. Rev. Applied 22, 054034 (2024).

37. J. Veen, A., Proutski, T. Karzig, D.I. Pikulin, R.M. Lutchyn, J. Nygård, P. Krogstrup, A. Geresdi, L.P. Kouwenhoven, J.D. Watson, Magnetic-field-dependent quasiparticle dynamics of nanowire single-Cooper-pair transistors. Phys. Rev. B 98, 174502 (2018).

38. D.J. Woerkom, A. Geresdi, L.P. Kouwenhoven, One minute parity lifetime of a NbTiN Cooper-pair transistor. Nat. Phys. 11(7), 547–550 (2015).

39. A. Bernard, Y. Peng, A. Kasumov, R. Deblock, M. Ferrier, F. Fortuna, V. T. Volkov, Yu. A. Kasumov, Y. Oreg, F. von Oppen, H. Bouchiat and S. Guéron, Long-lived Andreev states as evidence for protected hinge modes in a bismuth nanoring Josephson junction, Nat. Phys. 19, 358–364 (2023).

40. M. Endres, A. Kononov, H. S. Arachchige, J. Yan, D. Mandrus, K. Watanabe, T. Taniguchi, C. Schönenberger, Current–phase relation of a $WTe_2$ Josephson junction, Nano Lett., 23, 10, 4654–4659 (2023).

41. K. K. Likharev, Superconducting weak links, Rev. Mod. Phys. 51, 101 (1979).

42. R. S. Souto, M. Leijnse, and C. Schrade, Josephson diode effect in supercurrent interferometers, Phys. Rev. Lett. 129, 267702 (2022).

43. L.P. Kouwenhoven, C.M. Marcus, P.L. McEuen, S. Tarucha, R.M. Westervelt, N.S. Wingreen, In: L.L. Sohn, L.P. Kouwenhoven, G. Schön (eds.), Electron Transport in Quantum Dots, pp. 105–214. Springer, Dordrecht (1997).

44. J. Koch, T.M. Yu, J. Gambetta, A.A. Houck, D.I. Schuster, J. Majer, A. Blais, M.H. Devoret, S.M. Girvin, R.J. Schoelkopf, Charge-insensitive qubit design derived from the Cooper pair box. Phys. Rev. A 76, 042319 (2007).

45. S.M. Albrecht, A.P. Higginbotham, M. Madsen, F. Kuemmeth, T.S. Jespersen, J. Nygård, P. Krogstrup, C.M. Marcus, Exponential protection of zero modes in Majorana islands. Nature 531(7593), 206–209 (2016).

46. J. Shen, S. Heedt, F. Borsoi, B. Heck, S. Gazibegovic, R.L.M. Veld, D. Car, J.A. Logan, M. Pendharkar, S.J.J. Ramakers, G. Wang, D. Xu, D. Bouman, A. Geresdi, C.J. Palmstrøm, E.P.A.M. Bakkers, L.P. Kouwenhoven, Parity transitions in the superconducting ground state of hybrid InSb–Al Coulomb islands. Nat. Commun. 9(1), 4801 (2018).

47. D. Pan, M. Fu, X. Yu, X. Wang, L. Zhu, S. Nie, S. Wang, Q. Chen, P. Xiong, S. von Molnár, J. Zhao, Controlled synthesis of phase-pure InAs nanowires on Si(111) by diminishing the diameter to 10 nm. Nano Lett. 14, 1214-1220 (2014).

48. D. Pan, H. Song, S. Zhang, L. Liu, L. Wen, D. Liao, R. Zhuo, Z. Wang, Z. Zhang, S. Yang, J. Ying, W. Miao, R. Shang, H. Zhang, J. Zhao, In situ epitaxy of pure phase ultra-thin InAs-Al nanowires for quantum devices. Chinese Physics Letters 39(5), 058101 (2022).

49. X. Shi, Z. Dou, D. Pan, G. Li, Y. Li, A. Wang, Z. Zhang, X. Guo, X. Deng, B. Tong, Z. Lyu, P. Li, F. Qu, G. Liu, J. Zhao, J. Hu, L. Lu, J. Shen, Circuit-level-configurable zero-field superconducting diodes: a universal platform beyond intrinsic symmetry breaking, arXiv:2505.18330 (2025).

# Supplemental Information: Density of States (Gate) - Controlled Andreev Molecule and Sensor

Xiaofan Shi[1,4,†], Ziwei Dou[1,†,*], Guoan Li[1,4,†], Dong Pan[2,†], Yuxiao Song[1,4,†], Anqi Wang[1], Zhiyuan Zhang[1,4], Xingchen Guo[1,4], Xiao Deng[1,4], Yupeng Li[5], Bingbing Tong[1], Xiaohui Song[1], Zhaozheng Lyu[1], Peiling Li[1], Fanming Qu[1,4,6,7], Guangtong Liu[1,6,7], Jianhua Zhao[2,3,*], Li Lu[1,4,6,7,*], Jie Shen[1,6,*]

[1]Beijing National Laboratory for Condensed Matter Physics and Institute of Physics, Chinese Academy of Sciences, Beijing 100190, China

[2]State Key Laboratory of Semiconductor Physics and Chip Technologies, Institute of Semiconductors, Chinese Academy of Sciences, Beijing 100083, China

[3]National Key Laboratory of Spintronics, Hangzhou International Innovation Institute, Beihang University, Hangzhou 311115, China

[4]University of Chinese Academy of Sciences, Beijing 100049, China

[5]Hangzhou Key Laboratory of Quantum Matter, School of Physics, Hangzhou Normal University, Hangzhou 311121, China

[6]Songshan Lake Materials Laboratory, Dongguan, Guangdong 523808, China

[7]Hefei National Laboratory, Hefei 230088, China

* Corresponding authors. E-mails: ziweidou@iphy.ac.cn, jhzhao@semi.ac.cn, lilu@iphy.ac.cn, shenjie@iphy.ac.cn.

†These authors contributed equally to this work.

# Table of Content

## 1 Fast counter measurement

$I_C$ is commonly obtained by lock-in measurement where the device's differential resistance versus dc current bias and field $R(I_{DC}, B)$ is measured and $I_C$ is determined as the boundary between zero and non-zero $R$. Such method is usually slow since total measurement time for each $B$ is the lock-in demodulation time multiplies the number of points in $I_{DC}$ sweep. The fast counter measurement can obtain $I_C$ in a more efficient way, and is widely accepted in other JJ measurements[1-5]. In such setup (Supplementary Fig. 1), a triangular waveform of $I_{DC}$ is generated, and the amplified and filtered dc voltage of the device $V$ is sent to the digital counter. Meanwhile, a square wave (TTL) synchronized with $I_{DC}$ is also sent to the counter. The counter registers the time at the rising edge of $V$ (time for $I_{DC} = I_C$) and TTL waveforms (time for $I_{DC} = 0$). The time lapse between the two rising edges can be translated directly to $I_C$. To enhance the accuracy of $I_C$, the average from several repeated measurement is usually done at each $B$. The measurement time for each $B$ is then set by the period of the waveform $T$ multiplied by the number of repeated measurement, which can be made much faster than the lock-in technique. $T$ is chosen such that the results are independent of it.

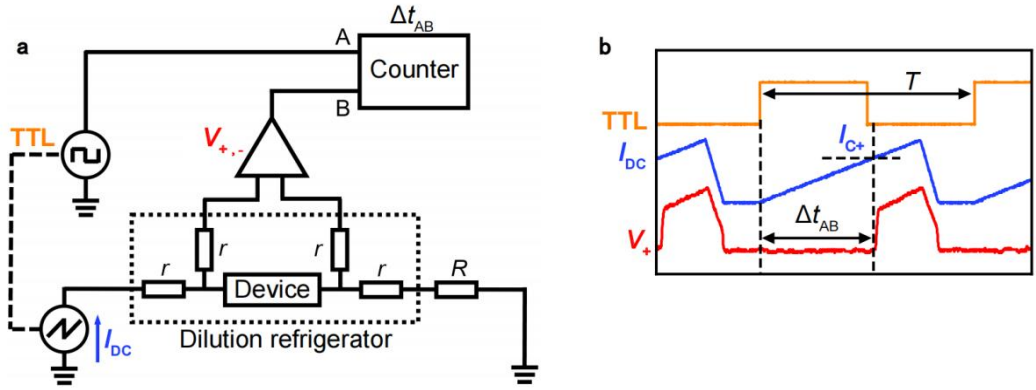

**Supplementary Fig. 1. Schematic for fast counter measurement setups: a,** a triangular wave $I_{DC}(t)$ (blue waveform in **b**) with period $T$. The device voltage $V$ (red waveform in **b**) has a sharp jump from 0 at $I_C$. $I_C$ is measured by the time lapse $\Delta t_{AB}$ registered in a digital counter, between the rising edges of $V$ and TTL (yellow waveform in **b**). **b,** An example of the measured time-domain waveforms.

## 2 Crosstalk between gates

Due to the unintentional capacitive coupling between various metallic gates in the device, changing the potential in one gate may nonlocally affect the other. Such crosstalk effect[6] widely observed in quantum devices may also cause a Fermi level shift in $JJ_S$ via TG1, TG2 or PG, which modifies $I_{C,S}$.

Here we characterize the strength of the crosstalk by first observing a weak $I_C(V_{SG})$ oscillation from $JJ_S$ (marked by dashed lines in Supplementary Figs. 2a-c), possibly due to unintentional quantum dots formed in the single $JJ_S$ constriction. Since such oscillation is sensitive to gate potential, it is helpful in determining the crosstalk. Indeed, the crosstalk from TG1, TG2 or PG to SG should result in a shift of the oscillations in the $V_{SG}$ direction. Supplementary Figs. 2a,b,c show $I_C(V_{SG})$ taken with wide range of $V_{TG1}$, $V_{TG2}$, $V_{PG}$ respectively. The $V_{SG}$ positions of the $I_C(V_{SG})$ oscillations are marked by dashed lines, and show negligible dependence on $V_{TG1}$, $V_{TG2}$, $V_{PG}$. These thus confirms the negligible crosstalk from TG1, TG2, PG to SG. Such negligible crosstalk is also consistent with the device structure where TG1, TG2, PG are separated by SG by the grounded middle superconducting contact.

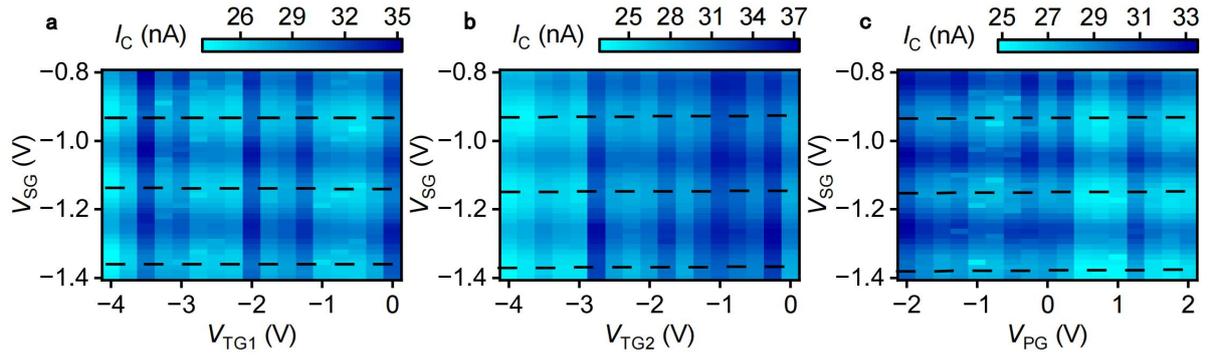

**Supplementary Fig. 2. Negligible crosstalk from $V_{TG1}$, $V_{TG2}$, $V_{PG}$ to $V_{SG}$: a,** $I_C(V_{SG})$ oscillations measured with various $V_{TG1}$. dashed lines: position of each oscillation peak versus $V_{TG1}$. No clear position shift is observed. **b, c,** Similar to **a** but with $V_{TG2}$ and $V_{PG}$, respectively.

## 3 Tight binding calculations of Andreev molecule

The DOS-controlled molecule device is modeled as an S-$N_1$-S-$N_S$-S, where S-$N_1$-S and S-$N_S$-S correspond to $JJ_1$ and $JJ_S$ in Fig. 2 respectively. Its Hamiltonian is written as:

$$\mathcal{H} = \begin{pmatrix} H - E_F & \Delta \\ \Delta^* & E_F - H^* \end{pmatrix} \quad (S1)$$

where $H$ is the Hamiltonian without superconductivity and $E_F$ is the Fermi level. The complex-valued pairing potential $\Delta = |\Delta|e^{i\varphi}$ incorporates the phase $\varphi$ of each S section. $|\Delta|$ is nonzero only in the S section. $\varphi = 0$ is set for the middle S section as the reference. $\varphi = -\varphi_1$ and $\varphi_S$ for the left and right S, respectively.

We model the junction by a 1d lattice using the KWANT package[7,8]. Discretizing (S1) thus leads to:

$$\mathcal{H} = \sum_{n=0}^{N}[E_n\boldsymbol{\sigma_z} + |\Delta|e^{i\varphi}\boldsymbol{\sigma_x}]|n\rangle\langle n| - \sum_{n=0}^{N-1}t_{n,n+1}\boldsymbol{\sigma_z}|n\rangle\langle n+1|$$

$$(S2)$$

where $n$ represents the n-th lattice and $N$ is the total lattice number, $\boldsymbol{\sigma}_{x,z}$ are the Pauli matrices operating on Nambu space. $t_{n,n+1} = t$ is assumed to be the constant hopping energy between the nearest neighbor lattices only. Higher order coupling is not considered. $E_n = 4t - E_F + U$ is the onsite potential of n-th lattice, where $U$ models the additional electrostatic potential in the normal regions due to gate tuning of $JJ_1$ and $JJ_S$. We set $E_F = 4t$ and $|\Delta| = 0.1t \ll E_F$ so that the retro-reflected electron-hole quasiparticles pairs in the conventional Andreev reflection are satisfied. The Andreev molecule effect is introduced by fixing the middle superconductor length $L = 7a$ shorter than the superconducting coherence length.

For each $U_{TG1}$, the Andreev spectrum of $JJ_S$ is thus calculated by fixing $\varphi_1$ of $JJ_1$ and diagonalizing (S2) for each $\varphi_S$ between 0 and $4\pi$ (with the periodicity of $2\pi$). The CPR of JJ$_S$ is calculated according to[8,10]:

$$I_{JJS}(\varphi) = \sum_m f_m \frac{\partial \epsilon_m}{\partial \varphi} \qquad (S3)$$

where $\epsilon_m$ is the m-th Andreev level and $f_m$ is the Fermi-Dirac distribution function at the level $\epsilon_m$. Similar calculation can also be done JJ$_1$ by fixing $\varphi_S$ and diagonalizing (S2) for each $\varphi_1$ between 0 and $4\pi$. Since we focus on the low temperature behaviors we set the temperature $T$ = 0.001t << $|\Delta|$. The critical current is then calculated from the CPR by $I_{C,S}$ = max$\{I_{JJS}(\varphi_S)\}$ and $I_{C,1}$ = max$\{I_{JJ1}(\varphi_1)\}$. For simplicity of this qualitative calculation, we adopt the convention that hopping energy t = 1, the lattice constant a = 1, the Boltzmann constant $k_B$ = 1, and the planck's constant $\hbar$ = 1.

To mimic the asymmetric SQUID in Fig. 2, $U_{SG}$ is set as 0 in Fig. 3 and JJ$_S$ thus has large DOS. We calculate the Andreev spectrum of JJ$_1$ (top panels) with $\varphi_S$ = 0 and JJ$_S$ (bottom panels) with $\varphi_1$ = 0 with a series of $U_{TG1}$ between -3.5t and -0.7t in Supplementary Figs. 3a-d. Similar to Fig. 3, the Andreev spectrum of JJ$_S$ with $\varphi_1$ = $\pi/4$ are plotted as the dashed lines to showcase the nonlocal phase control of JJ$_S$ ABSs by $\varphi_1$ due to wavefunction hybridization. The double arrows at $\varphi_S$ = $\pi/2$ reflect the wavefunction transmission, as explained in the main text. From $U_{TG1}$ = -3.5t to -0.7t, the Fermi level increases in N$_1$, and the ABSs of JJ$_1$ start to appear. Meanwhile, the ABSs of JJ$_S$ are also nonlocally affected by $U_{TG1}$: 1. Wavefunction hybridization becomes more pronounced, indicated by the shift of the ABSs of JJ$_S$ between $\varphi_1$ = 0 (solid lines) and $\varphi_1$ = $\pi/4$ (dashed lines). 2. Wavefunction transmission enhances, manifested as the reduced gap at $\varphi_S$ = $\pi/2$ (double arrows). Both mechanisms become prominent with higher DOS in N$_1$.

We note that the nonlocal control of $I_{C,S}$ by $U_{TG1}$ shown in Figs. 3b-d remains qualitatively the same with other values of $\varphi_1$ and $\varphi_S$, as shown in Supplementary Fig. 4 with

$\varphi_1 = \pi/4$ and $\varphi_S = \pi/2$.

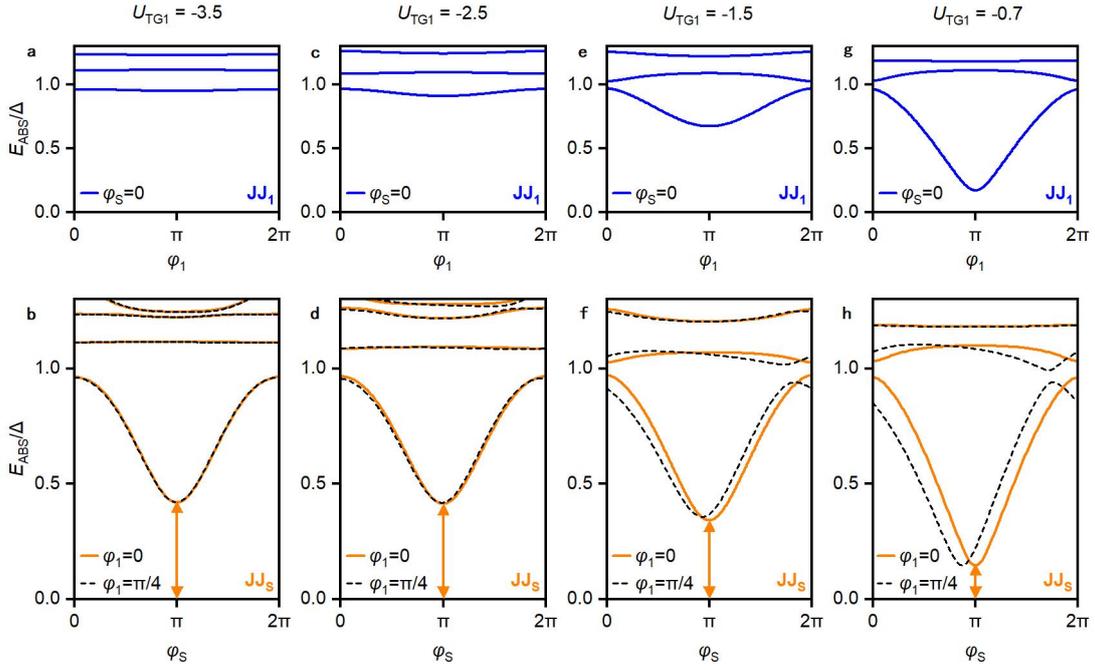

**Supplementary Fig. 3, Andreev levels of JJ$_S$ with varying $U_{TG1}$: a-d,** Calculated Andreev levels of JJ$_1$ versus $\varphi_1$ with $\varphi_S = 0$ (blue lines, top panels) and JJ$_S$ versus $\varphi_S$ with $\varphi_1 = 0$ (orange lines, bottom panels). $U_{TG1}$ = -3.5, -2.5, -1.5, -0.7 respectively. Dashed lines correspond to Andreev levels of JJ$_S$ versus $\varphi_S$ with $\varphi_1 = \pi/4$. The shifts between $\varphi_1 = 0$ and $\pi/4$ levels reflect the wavefunction hybridization. The gap at $\varphi_S = \pi$ (double arrows) reflects the wavefunction transmission. Both mechanisms are nonlocally tuned by $U_{TG1}$.

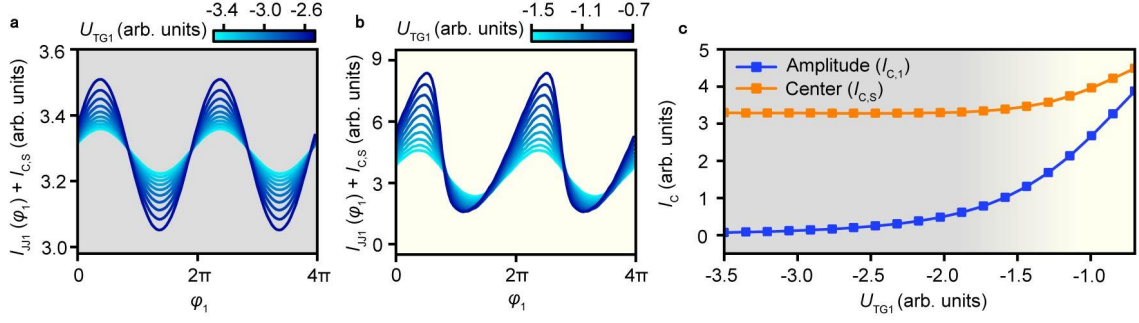

**Supplementary Fig. 4, Similar nonlocal response of $I_{C,S}$ of $U_{TG1}$ for non-zero phases: a,b,** Simulated $I_{JJ1}(\varphi_1)+I_{C,S}$ with different $U_{TG1}$ for non-molecule (**a**) and molecule regimes (**b**), similar to Figs. 2c,d. $\varphi_1$ across JJ$_1$ grows proportionally with the SQUID flux $\Phi$. $I_{JJ1}(\varphi_1)$ is calculated with $\varphi_S = \pi/2$; $I_{C,S}$ is calculated with $\varphi_1 = \pi/4$. **c,** Extracted $I_{C,1}$ and $I_{C,S}$ versus $U_{TG1}$ also showing similar features as Figs. 3d.

# 4 Possible Andreev molecule effects in JJ$_2$ and JJ$_1$ inferred by the shape of SQUID oscillations

In the main text, we focus on the Andreev molecule effects of JJ$_S$ via the nonlocal control of $I_{C,S}$ by TG1,2 and PG. With the SQUID readout technique, the similar Andreev molecule effects in the locally tuned JJ$_2$ and JJ$_1$ can be inferred by the shape of the SQUID oscillations $\Delta I_C(B)$, since in the asymmetric SQUID, $\Delta I_C(B)$ reflects the shape of the CPR of the weakest junctions[9].

We first focus on the case of JJ$_2$ in Fig. 4. Here, the SQUID oscillation shape $\Delta I_C(B)$ reflects the the CPR in the weakest JJ$_2$. The conventional CPR of a JJ can be approximated as $I(\varphi) = \sum_n t^n \frac{(-1)^{n+1}}{n} \sin(n\varphi)$, where $t$ is the JJ transmission with $0 \leq t \leq 1$. The second and higher harmonic (that is, $\sin(n\varphi)$ terms with n $\geq$ 2) are present when $t$ is significantly larger than 0 at higher DOS[41]. In such conventional CPR without time-reversal symmetry breaking (TRSB), $I(-\varphi)=-I(\varphi)$ and the CPR thus has a centro-symmetric shape. However, when the wavefunction hybridization happens due to the extended wavefunction of JJ$_1$ into JJ$_2$, the anomalous phase shift of the JJ$_2$ ABSs may be present due to the Andreev molecule effects[12]. The centro-symmetric shape of the CPR and the measured $\Delta I_C(B)$ will be further broken when the transmission of JJ$_2$ is high enough with sizeable second harmonic[12]. Indeed, Supplementary Fig. 5a directly shows a typical case in the molecule regime with $V_{TG2}$ = 0 V. The measured SQUID oscillation minus its respective center value (noted as $\Delta I_C$) is plotted with the same data under the centro-symmetric operation (noted as the "inverse" curve). By overlaying the original and the inverse curves, we directly see a visible difference (highlighted by the red arrow) in the molecule regime, similar to the non-centro-symmetric shape of the CPRs observed in previous type I Andreev molecule works[11,13]. We emphasize that such non-centro-symmetric shape of the CPR of JJ$_2$ appears around zero flux of the SQUID and cannot be associated with flux-induced TRSB effect around $\Phi_0/2$[12]. Therefore, the non-centro-symmetric and thus the TRSB CPR of $JJ_2$ indicates a wavefunction extension of $JJ_1$ into $JJ_2$ and thus supports scenario (A) more than scenario (B) in Fig. 4.

In Supplementary Fig. 5b, similar non-centro-symmetric CPR of $JJ_1$ is also observed in the case of Fig. 2, in the molecule regime with $V_{TG1}$ = 3.0 V, possibly due to the wavefunction extension of $JJ_S$ into $JJ_1$.

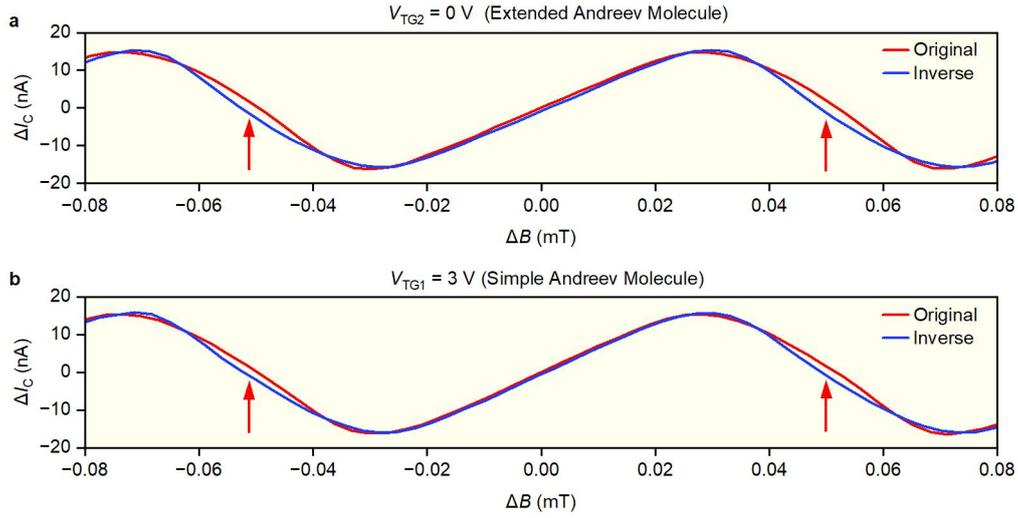

**Supplementary Fig. 5, Non-centro-symmetric SQUID oscillations in the molecule regime:** **a,** Measured SQUID oscillation shape with its center value removed ($\Delta I_C(B)$) reflects the CPR of JJ$_2$, showing non-centro-symmetric shape (highlighted by the red arrow) in the molecule regime of the extended Andreev molecule. The data is reproduced from Fig. 4d with $V_{TG2} = 0$ V. **b,** Similar non-centro-symmetric $\Delta I_C(B)$ and the CPR of JJ$_1$ (highlighted by the red arrow) in the molecule regime of the simple Andreev molecule. The data is reproduced from Fig. 2d with $V_{TG1} = 3.0$ V.

**References**

1. G.P. Mazur, N. van Loo, D. van Driel, J.-Y. Wang, G. Badawy, S. Gazibegovic, E.P.A.M. Bakkers, L.P. Kouwenhoven, Gate-tunable Josephson diode, Phys. Rev. Applied 22, 054034 (2024).

2. J. Veen, A., Proutski, T. Karzig, D.I. Pikulin, R.M. Lutchyn, J. Nygård, P. Krogstrup, A. Geresdi, L.P. Kouwenhoven, J.D. Watson, Magnetic-field-dependent quasiparticle dynamics of nanowire single-Cooper-pair transistors. Phys. Rev. B 98, 174502 (2018).

3. D.J. Woerkom, A. Geresdi, L.P. Kouwenhoven, One minute parity lifetime of a NbTiN Cooper-pair transistor. Nat. Phys. 11(7), 547–550 (2015).

4. A. Bernard, Y. Peng, A. Kasumov, R. Deblock, M. Ferrier, F. Fortuna, V. T. Volkov, Yu. A. Kasumov, Y. Oreg, F. von Oppen, H. Bouchiat and S. Guéron, Long-lived Andreev states as evidence for protected hinge modes in a bismuth nanoring Josephson junction, Nat. Phys. 19, 358–364 (2023).

5. M. Endres, A. Kononov, H. S. Arachchige, J. Yan, D. Mandrus, K. Watanabe, T. Taniguchi, C. Schönenberger, Current–phase relation of a WTe2 Josephson junction, Nano Lett., 23, 10, 4654–4659 (2023).

6. L.P. Kouwenhoven, C.M. Marcus, P.L. McEuen, S. Tarucha, R.M. Westervelt, N.S. Wingreen, In: L.L. Sohn, L.P. Kouwenhoven, G. Schön (eds.), Electron Transport in Quantum Dots, pp. 105–214. Springer, Dordrecht (1997).

7. C. W. Groth, M. Wimmer, A. R. Akhmerov and X. Waintal, Kwant: a software package for quantum transport, New J. Phys., 16, 063065 (2014).

8. M. Ferrier, B. Dassonneville, S. Guéron, and H. Bouchiat, Phase-dependent Andreev spectrum in a diffusive SNS junction: Static and dynamic current response. Phys. Rev. B,

88(17): 174505 (2013).

9. M. Tinkham: Introduction to Superconductivity. International series in pure and applied physics. McGraw Hill, New York (1996).

10. K. K. Likharev, Superconducting weak links, Rev. Mod. Phys. 51, 101 (1979).

11. J.-D. Pillet, S. Annabi, A. Peugeot, H. Riechert, E. Arrighi, J. Griesmar, and L. Bretheau, Josephson diode effect in Andreev molecules, Phys. Rev. Research 5, 033199 (2023).

12. R. S. Souto, M. Leijnse, and C. Schrade, Josephson diode effect in supercurrent interferometers, Phys. Rev. Lett. 129, 267702 (2022).

13. S. Matsuo, T. Imoto, T. Yokoyama, Y. Sato, T. Lindemann, S. Gronin, G. C. Gardner, M. J. Manfra, S. Tarucha, Phase engineering of anomalous Josephson effect derived from Andreev molecules, Sci. Adv. 9, eadj3698 (2023).